# Heisenberg-like critical behavior in the quasi-two-dimensional metallic ferromagnet $LaCrSb_3$*

MAO Qianhui[1, *], DENG Haotian[1], CHEN Bin[2], YANG Jinhu[2]

1. College of Science, Henan University of Engineering, Zhengzhou 451191, China
2. Department of Physics, Hangzhou Normal University, Hangzhou 311121, China

**Abstract** $LaCrSb_3$ is a material exhibiting both quasi-two-dimensional spin fluctuations and three-dimensional magnetic interaction characteristics. By measuring the isothermal magnetization of single-crystals and conducting a systematic critical behavior analysis, we clarify the critical properties of its ferromagnetic phase transition and the intrinsic magnetic interaction mechanism. Based on high-precision isothermal magnetization data measured in the vicinity of the critical point, the Curie temperature for the ferromagnetic-paramagnetic phase transition is determined to be $T_C = 126$ K, with the critical exponents obtained as $\beta = 0.376$, $\gamma = 1.417$ and $\delta = 4.76$ via the self-consistent iterative method based on the Arrott-Noakes equation. The reliability of these critical exponents is verified by the Widom scaling law, the magnetic state scaling equation and other analyses. A comparison with theoretical models demonstrates that the critical behavior of the magnetic phase transition in this system basically belongs to the universality class of the three-dimensional Heisenberg model. This conclusion is further confirmed by the distance-dependent decay behavior of the exchange interaction $J(r)$, revealing the dominant role of isotropic direct exchange interactions in this system. Finally, drawing on research findings of other quasi-two-dimensional magnetic materials, this work proposes that $LaCrSb_3$ may exhibit finite-temperature magnetic order in the two-dimensional limit, thereby possessing important theoretical research significance and promising practical application prospects.

---

# 1 Introduction

Low-dimensional magnetic materials have long been a research focus in condensed matter physics and materials science. This interest stems from their exotic physical properties arising from strong spin fluctuations and their potential applications in next-generation high-speed information devices [1-8]. Specifically, the high-temperature superconductivity emerging near antiferromagnetic instability in cuprates, nickelates, and iron arsenides/selenides provides key insights for exploring unconventional superconductors [5-8]. Beyond this macroscopic quantum effect, low dimensionality plays a significant role in the formation of charge/spin density waves, giant magnetoresistance, and unconventional quantum critical points [9-11]. In terms of applied research, the discovery of tunable two-dimensional magnetism and excellent physical properties in materials such as $Fe_3GeTe_2$ [1], $CrI_3$ [12], and $CrGeTe_3$ [13] has greatly advanced the fields of spintronics and magnonics. Despite significant progress in studying low-dimensional magnetic materials, the exploration of novel low-dimensional magnetic systems remains a core direction for researching spin-related physical effects and developing devices.

The intermetallic antimonide $LaCrSb_3$ has attracted widespread attention in recent years due to its unique crystal and magnetic properties [14-28]. This material belongs to the orthorhombic crystal system (space group *Pbcm*, No. 57). In its crystal structure, $CrSb_6$ octahedra connect via edge-sharing and face-sharing along the *b* and *c* axes, respectively. The corrugated layers of $CrSb_6$ octahedra are separated by $La^{3+}$ ions and nearly square-planar Sb atom layers, forming a typical layered structure [29]. The phase transition temperature $T_C$ from the paramagnetic to the ferromagnetic state in $LaCrSb_3$ exhibits robustness, showing weak modulation by doping and pressure. Its phase transition temperature ranges from 125 to 147 K, with a saturation magnetic moment of $0.8 \leq P_S \leq 1.7$ $\mu_B$/Cr. Neutron diffraction experiments indicate that both ferromagnetic and canted antiferromagnetic sublattices coexist in the system when the temperature is below $T_C$ = 126 K. Furthermore, at 95 K, the spontaneous magnetic moment of the system undergoes a crystal axis reorientation, shifting from the crystal *b* axis to the *c* axis [20]. This unique magnetic structure was once considered an ideal platform for finely tuning quantum tricritical points. However, the doped system $LaCr_{1-x}Fe_xSb_3$ did not exhibit the expected quantum tricritical point due to a first-order canted magnetic phase transition [28]. Additionally, previous studies reported that the anomalous Hall conductivity of this material reaches as high as 1250 $\Omega^{-1}\cdot cm^{-1}$ at 2 K, comparable to that of the topological Weyl semimetal $Co_3Sn_2S_2$. This phenomenon is attributed to topologically nontrivial band inversion and flat band opening effects near the Fermi level [26]. Recently, we fitted and analyzed the magnetization data of $LaCrSb_3$ to extract spin fluctuation parameters based on Yoshinori Takahashi's theory. We found that its quantum spin fluctuations exhibit strong two-dimensionality. Moreover, in the Deguchi-Takahashi plot, the position of $LaCrSb_3$ is very close to that of $Fe_3GeTe_2$, suggesting that this material may have potential for spintronic applications [29]. Critical behavior analysis of bulk $LaCrSb_3$ single crystals grown by the flux method revealed that magnetic interactions possess three-dimensional, long-range characteristics. Comparison with theoretical models indicates that the magnetic Hamiltonian of this system cannot be classified into a single universality class [27].

The effective Hamiltonian of a magnetic system determines the dimensionality and intensity of spin fluctuations, thereby influencing the formation of magnetic order and spin-related transport behavior. $LaCrSb_3$ exhibits a unique low-dimensional structure, spin fluctuations similar to those in $Fe_3GeTe_2$, and nontrivial band topology, making it promising for next-generation spintronic devices. Therefore, critical questions naturally arise, such as how to reconcile the conflict between its quasi-two-dimensional spin fluctuations and three-dimensional critical exponents, and whether the Hamiltonian governing this magnetic system favors the formation of magnetic order in the two-dimensional limit. Consequently, further confirmation of the intrinsic magnetic interactions in this system becomes crucial. In this study, we obtained high-quality single crystals using stoichiometric elements via the melt-cooling method. Through detailed isothermal magnetization measurements and data fitting analysis, we self-consistently determined the critical exponents near the ferromagnetic phase transition point of this system. Using renormalization group conclusions, we derived the relationship describing how the coupling strength of magnetic moments decays with distance. The results show that the critical exponents of $LaCrSb_3$ are basically consistent with the predictions of the Heisenberg model. Meanwhile, the magnetic interaction decays rapidly with distance, indicating that short-range direct exchange interactions dominate in this system.

# 2 Experimental Methods

$LaCrSb_3$ single crystals used in the experiment were obtained via the direct melt-cooling method. High-purity La, Cr, and Sb elemental powders were thoroughly mixed with a stoichiometric molar ratio of La∶Cr∶Sb = 1∶1∶3 inside an argon-filled glove box. The mixture was then loaded into an alumina crucible placed inside a quartz tube and sealed under vacuum conditions ($<1.0\times10^{-2}$ Pa). The sealed quartz ampoule was heated inside a muffle furnace to 1373 K and held at this temperature for 6 h to guarantee complete melting and reaction of the raw materials. Subsequently, the temperature was slowly reduced to 773 K at a rate of 5 K/h, after which the power was turned off to allow the reactants to cool with the furnace. The typical size of the obtained crystals was approximately $a\times b\times c \approx 0.2\times0.8\times0.8$ mm$^3$. The single-crystal structure was determined using a Bruker D8 Quest diffractometer. The morphology and elemental ratio of the samples were determined using a scanning electron microscope (SEM, model Quanta 250) equipped with an energy dispersive X-ray spectroscopy (EDXS) detector. DC isothermal magnetization curves were measured using a vibrating sample magnetometer (VSM) on a magnetic testing system (Quantum Design MPMS-3) at the Department of Physics, Hangzhou Normal University. To guarantee the accuracy of data analysis, the temperature range near the critical point was set to 115-140 K with a step size of $\Delta T$ = 1 K. The magnetic field was set to a continuous scanning mode with equal intervals: a step size of $\Delta H$ = 100 Oe in the 0-2000 Oe range, and a step size of 1000 Oe in the 2000-50000 Oe range. During the testing process, an oscillating field decay mode was used to reduce the magnetic field strength to zero for each isotherm to eliminate the influence of remanent magnetization.

# 3 Results and Discussion

The characterization results of the single-crystal structure and composition are shown in Figure 1. The left inset in the figure displays the full diffraction spectrum of the (*h*0*l*) plane in reciprocal space. The diffraction spots are bright and regular in shape, with no significant blurring or trailing, indicating the high quality of the obtained single crystals. The resolved space group is orthorhombic (*Pbcm*, No. 57), with lattice constants $a$ = 13.274 Å, $b$ = 6.232 Å, and $c$ = 6.114 Å, which are consistent with previously reported values. The right inset shows that the characteristic length of the single crystal is 1 mm, and the cleavage planes are clearly visible, consistent with its quasi-two-dimensional microscopic structure. The main panel displays the EDXS results, which show characteristic peaks corresponding to La, Cr, and Sb elements, except for a small background signal from C atoms near zero energy. Quantitative calculation yields an atomic ratio of La:Cr:Sb = 0.97:0.99:3.04, which agrees very well with the nominal composition.

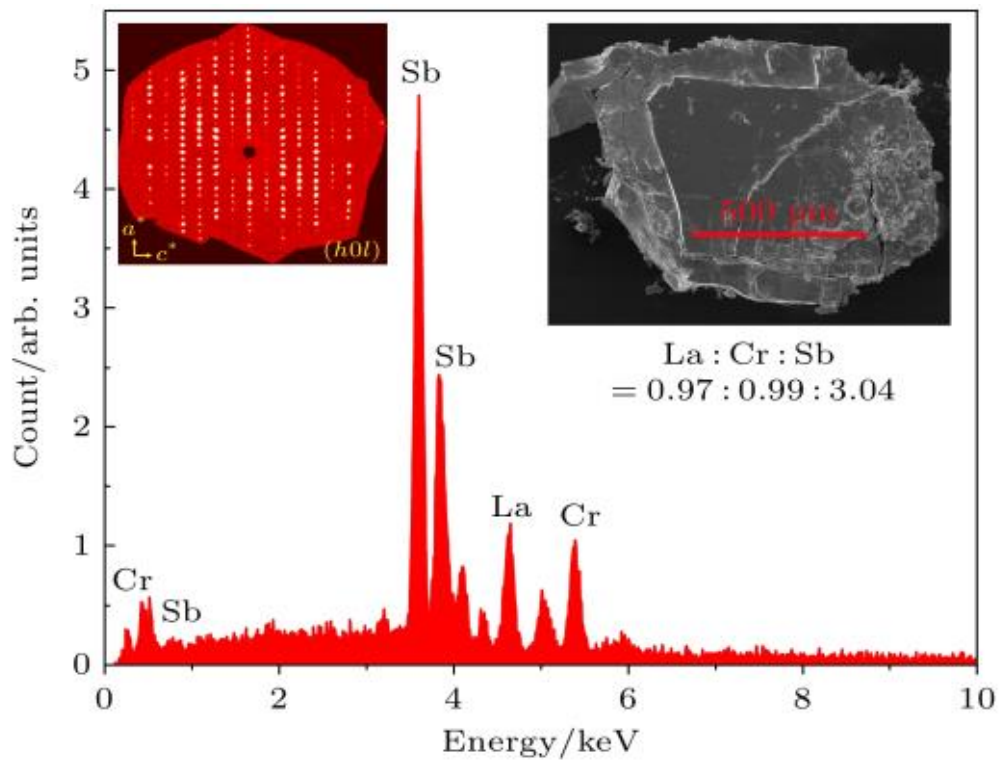

**Fig. 1 EDXS spectrum of $LaCrSb_3$ (main panel), single-crystal diffraction pattern in reciprocal space (left inset) and morphology (right inset).**

Figure 2(a) displays the isothermal magnetization $M$ as a function of magnetic field strength $H$, denoted as $M(H)$, measured within the temperature range of 2-300 K. In the low-temperature region, $M$ rises rapidly with increasing $H$ and quickly approaches saturation, indicating that the sample is in a ferromagnetic state. As the temperature increases, the saturation behavior gradually becomes less distinct and disappears near the Curie temperature $T_C$ = 126 K. It is replaced by a general morphology resembling a Brillouin function. Upon further heating, $M(H)$ exhibits linear behavior, corresponding to the paramagnetic region of the Brillouin function under high-temperature and low-field conditions. Across the entire temperature range, the magnetization changes continuously and smoothly with the applied magnetic field. No first-order phase transition features, such as abrupt magnetization jumps or metamagnetic transitions, were observed. For systems with long-range ferromagnetic interactions, the paramagnetic-to-ferromagnetic phase transition can typically be described using the Landau mean-field theory approximation. Its characteristics can be verified via Arrott plots ($M^2$ versus $H/M$), as shown in Figure 2(b). If mean-field theory applies, the isotherms at different temperatures should appear nearly parallel, and the isotherm at the critical temperature $T = T_C$ should pass through the origin. Based on this principle, linear extrapolation of the isotherms in the high-field region toward the vertical and horizontal axes yields the spontaneous magnetization $M_S(T, 0)$ and the inverse initial susceptibility $\chi_0^{-1}(T)$, respectively. These values provide experimental basis for subsequent critical exponent extraction and scaling analysis. The Arrott plot also indicates the type of phase transition according to the Banerjee criterion: a negative slope implies a first-order phase transition, whereas its absence indicates a continuous phase transition. As observed in the figure, the slopes of $M^2$ versus $H/M$ are all positive, indicating that the system undergoes a continuous phase transition. This finding corroborates our discussion of Figure 2(a). Furthermore, the Arrott plots exhibit non-zero curvature across all temperatures and magnetic field ranges, suggesting that mean-field theory provides only a rough description of the magnetism in $LaCrSb_3$.

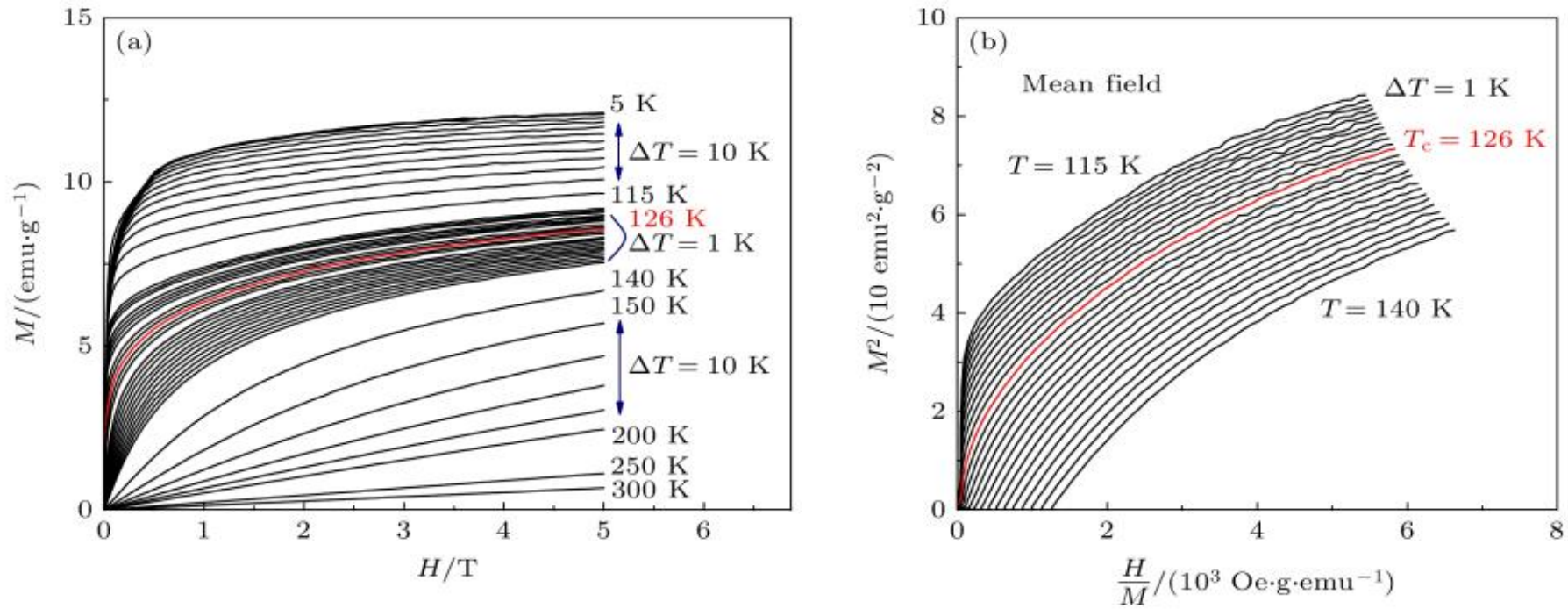

**Fig. 2 (a) Isothermal magnetization curves at 2–300 K; (b) Arrott plot of panel (a) around critical point.**

A more universal and reliable method for analyzing critical behavior is to employ the Arrott-Noakes equation[30-32]:

$$(H/M)^{1/\gamma} = A\varepsilon + BM^{1/\beta}, \qquad (1)$$

Here, $\varepsilon = (T - T_C)/T_C$ denotes the reduced temperature, while $\beta$ and $\gamma$ represent the critical exponents describing $M_S(T, 0)$ and $\chi_0^{-1}(T)$, respectively. This equation encompasses different universality classes. The Arrott plot corresponding to mean-field theory can be viewed as a special case of the equation of state with $\beta = 0.5$ and $\gamma = 1$. Figure 3 presents Arrott-Noakes plots constructed based on four different theoretical models: the two-dimensional Ising model ($\beta = 0.345$, $\gamma = 1.316$, Figure 3(a)), the three-dimensional Ising model ($\beta = 0.325$, $\gamma = 1.24$, Figure 3(b)), the three-dimensional Heisenberg model ($\beta = 0.365$, $\gamma = 1.386$, Figure 3(c)), and the tricritical mean-field model ($\beta = 0.25$, $\gamma = 1$, Figure 3(d)). Although $LaCrSb_3$ exhibits strong two-dimensional spin fluctuations[29], the two-dimensional Ising model shown in Figure 3(a) fails to yield the spontaneous magnetization below $T_C$. Instead, it produces negative, unphysical results. The other three models display better parallel linear behavior. Furthermore, the isotherm at $T = 126$ K extrapolates to near the origin on the coordinate axes. This indicates that the Curie temperature $T_C$ of the system is close to 126 K. However, the parallelism of the isotherms alone is insufficient to definitively determine which model best fits the experimental data. Therefore, we introduce the normalized slope ($N_S$) as a quantitative criterion[33], defined as

$$N_S \equiv \frac{\mathrm{d}(M^{1/\beta})}{\mathrm{d}((H/M)^{1/\gamma})}\Big|_T \Big/ \frac{\mathrm{d}(M^{1/\beta})}{\mathrm{d}((H/M)^{1/\gamma})}\Big|_{T_C}. \qquad (2)$$

This metric reflects the similarity between isotherms at different temperatures and the isotherm at the critical temperature $T_C$. Ideally, if a theoretical model accurately describes the critical behavior of the system, its corresponding $N_S$ value should approach 1 near the critical temperature $T_C$ and remain insensitive to temperature variations.

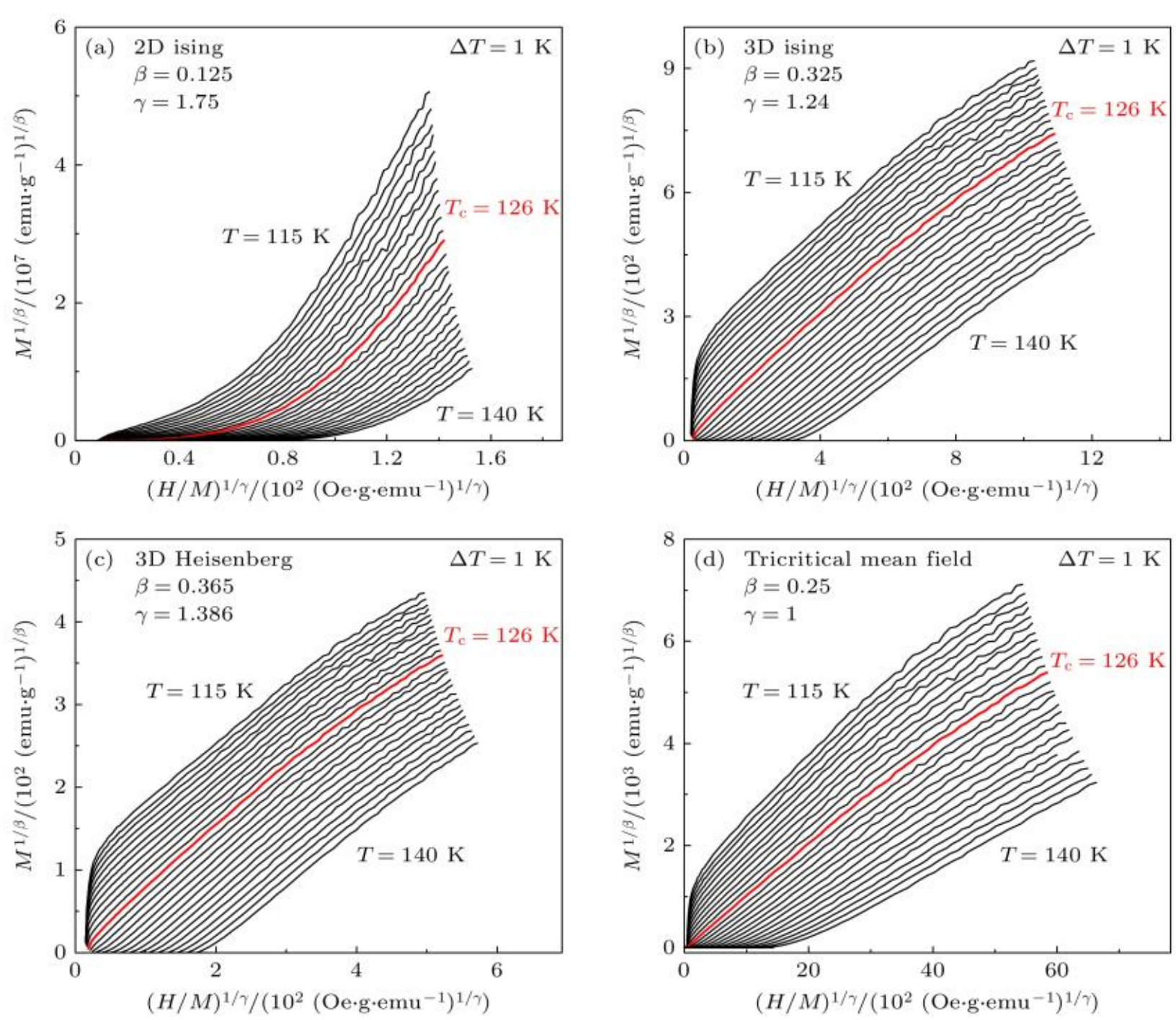

**Fig. 3 The isotherms plotted as $M^{1/\beta}$ versus $(H/M)^{1/\gamma}$ with different models: (a) 2D Ising model; (b) 3D Ising model; (c) 3D Heisenberg model; (d) tricritical mean-field model.**

Figure 4 clearly demonstrates that among the compared models, the three-dimensional Heisenberg model yields an $N_S$ value closest to 1. This indicates that this model is most suitable for determining $M_S(T, 0)$, $\chi_0^{-1}(T)$, and $T_C$. To ensure the self-consistency of this method, one can fit the corresponding $\beta$ and $\gamma$ values once a series of $M_S(T, 0)$ and $\chi_0^{-1}(T)$ are obtained. This step verifies whether these parameters are consistent with the exponents used (convergence criterion). If the result is negative, new $\beta$ and $\gamma$ values must be substituted into the Arrott-Noakes equation, and the process repeated until a positive determination is achieved. It is important to note that self-consistency is guaranteed solely by passing the convergence test; an optimal model is not strictly required. However, selecting the best model initially is key to ensuring rapid convergence.

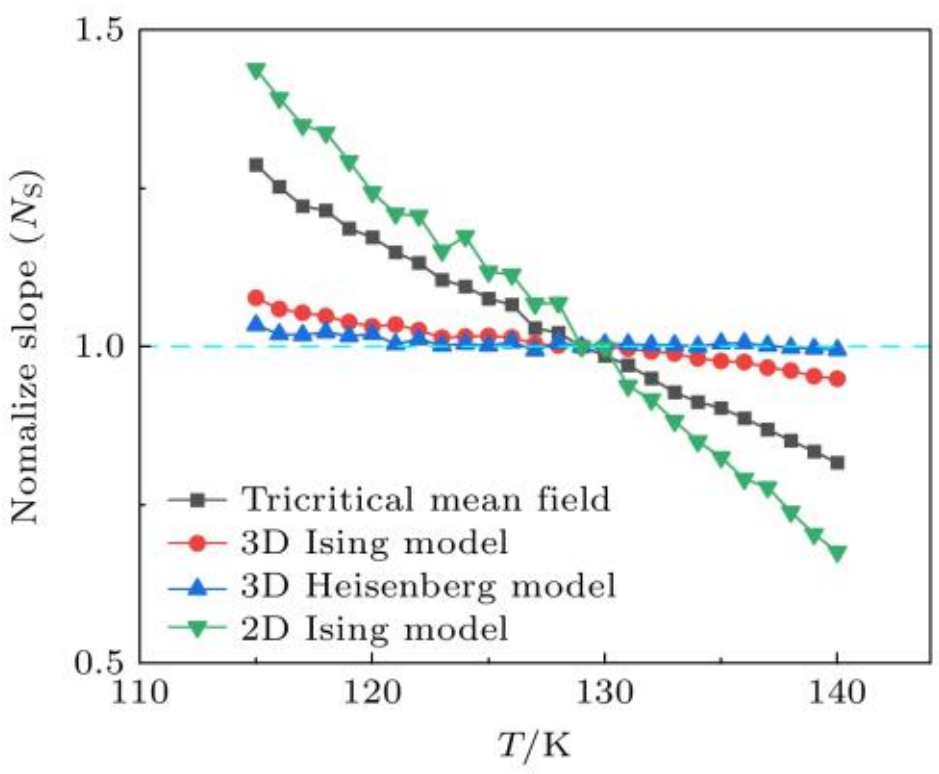

**Fig. 4 The normalized slope of different theoretical models as a function of temperature.**

Figure 5(a) presents the iteratively converged $M_S(T, 0)$ and $\chi_0^{-1}(T)$. Their temperature dependence facilitates the determination of the critical exponents of the system. According to the theory of critical phenomena, the following power-law relations hold within the critical region for $M_S(T, 0)$, $\chi_0^{-1}(T)$, and the critical isotherm $M(T_C)$:

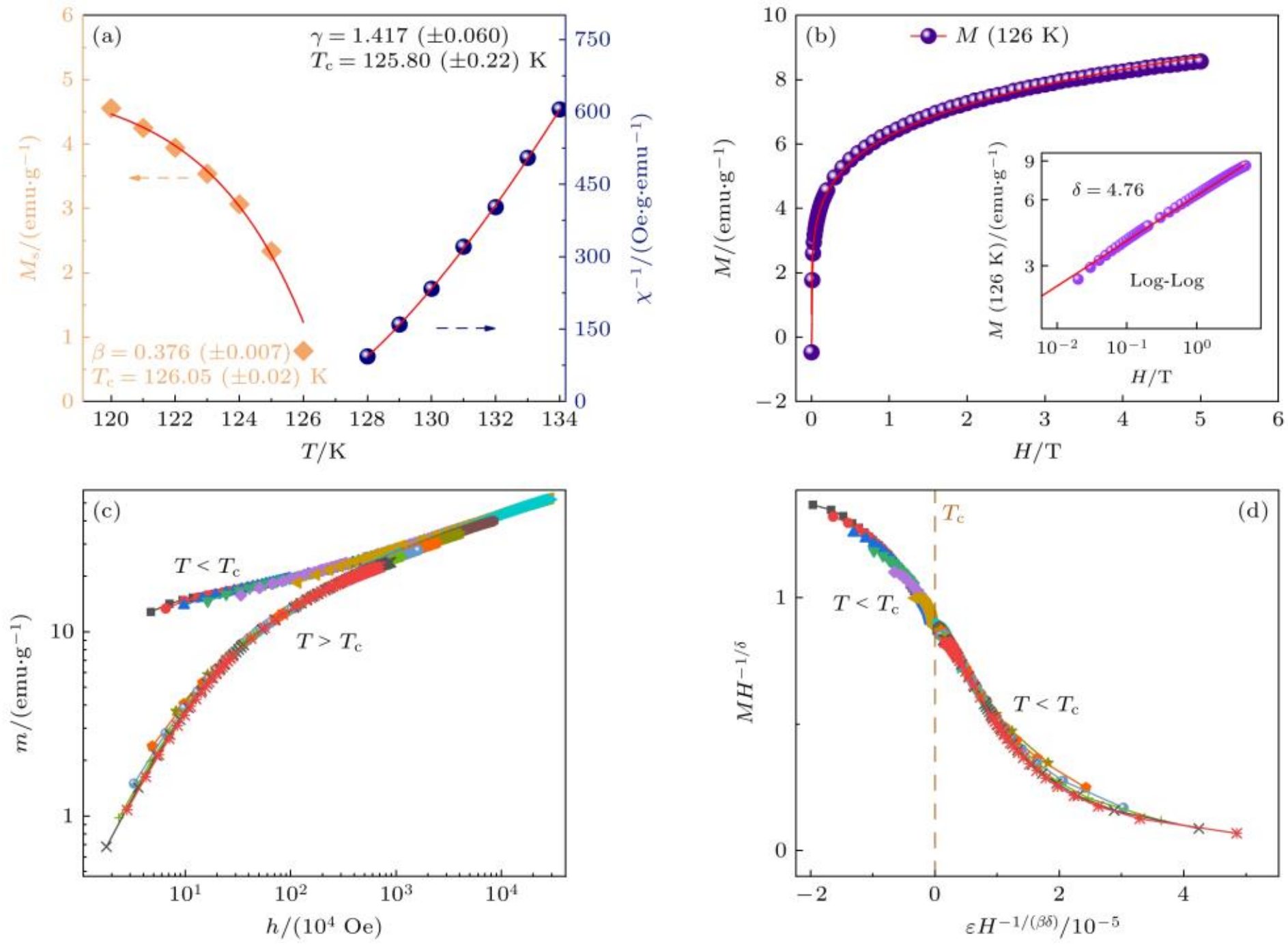

**Fig. 5 (a) Temperature dependence of $M_S$ ($T$, 0) and $\chi_0^{-1}(T)$, and their curves fitted using Eqs. (3) and (4); (b) isothermal magnetization curve at 126 K, insets shows log-log plots of the main figure and their linear fits; (c), (d) renormalized isothermal magnetization curves calculated using Eqs. (6) and (7), confirming their universal behavior.**

$$M_s(T, 0) = M_0(-\varepsilon)^{\beta}, \varepsilon < 0, T < T_C, \quad (3)$$

$$\chi_0^{-1}(T, 0) = \frac{h_0}{M_0}\varepsilon^{\gamma}, \varepsilon > 0, T > T_C, \quad (4)$$

$$M = DH^{1/\delta}, \varepsilon = 0, T = T_C, \quad (5)$$

Here, $\varepsilon$, $\beta$, and $\gamma$ are defined as in Eq. (1), while $M_0$, $(h_0/M_0)$, and $D$ denote the corresponding critical amplitudes. The exponent $\delta$ is another independent critical exponent that describes the behavior of magnetization under an external magnetic field at the critical temperature $T_C$. We fitted $M_S(T, 0)$ and $\chi_0^{-1}(T)$ using Eqs. (3) and (4), treating the critical exponents, critical amplitudes, and $T_C$ as free parameters. This yielded critical exponents of $\beta = 0.376$ (corresponding to $T_C = 126.05$ K) and $\gamma = 1.417$ (corresponding to $T_C = 125.80$ K). Notably, the $T_C$ values obtained from fitting different physical quantities are highly consistent with each other. They also agree well with the critical temperature derived from Arrott-Noakes plot analysis, indicating high reliability for the extracted critical exponents and $T_C$. Therefore, based on these results, we determine the Curie temperature of this system to be $T_C$ = 126 K.

Figure 5(b) and the inset display the magnetization curves $M(H, T_C)$ and their corresponding log-log representations, respectively. According to Eq. (5), we can directly extract the critical exponent $\delta$ by performing a linear fit on the log-log plot, yielding $\delta = 4.76$. Based on the scaling hypothesis, the Gibbs potential of a magnetic system is a homogeneous function of $H$ and $\varepsilon$. Consequently, only two critical exponents are independent. The exponents $\beta$, $\gamma$, and $\delta$ are related via the Widom scaling law: $\delta = 1+\beta/\gamma$. Substituting the $\beta$ and $\gamma$ values obtained from the Arrott-Noakes equation fit into the Widom scaling relation yields $\delta = 4.77$. This value is very close to the result $\delta = 4.76$ derived from the analysis of the critical magnetization curve $M(H, T_C)$.

Furthermore, we can verify the reliability of the critical exponents through scaling analysis[34]. Based on the assumption of thermodynamic function homogeneity, the critical exponents in Eqs. (3)—(5) can be unified and reformulated into magnetic equation of state expressions that describe magnetic behavior near $T_C$. Two common forms are

$$M(H,\varepsilon) = \varepsilon^{\beta} f_{\pm}\left(\frac{H}{\varepsilon^{\beta+\gamma}}\right), \qquad (6)$$

$$H = M^{\delta} f(\varepsilon M^{-1/\beta}), \qquad (7)$$

Here, $f_+$ and $f_-$ denote the scaling functions for $T > T_C$ and $T<T_C$, respectively. By introducing the dimensionless variables $m \equiv \varepsilon^{-\beta} M(H,\varepsilon)$ and $h \equiv H\varepsilon^{-(\beta+\gamma)}$, the scaling relation in Eq. (6) predicts that $m$ should be a two-branched function of $h$. According to Eq. (7), when the experimental data are rescaled according to the relationship between $MH^{-1/\delta}$ and $\varepsilon H^{-1/(\beta\delta)}$, all data points should collapse onto a universal curve and converge to the origin of the horizontal axis at $T = T_C$. As shown in Fig. 5(c) and (d), the normalized magnetization data collapse well onto the upper and lower branches and a single curve. This further validates the self-consistency and reliability of the obtained critical exponents.

Table 1 compares the critical exponents obtained in this study with predictions from different theoretical models and previous reports. These critical exponents basically belong to the three-dimensional Heisenberg universality class. This indicates that direct exchange plays a dominant role in the magnetic interactions of $LaCrSb_3$.

**Table 1** Comparison of the critical exponents with theoretical models and literature results.

| Experimental results or theoretical models | References | $\beta$ | $\gamma$ | $\delta$ |
|---|---|---|---|---|
| $LaCrSb_3$ | Text | 0.376 | 1.417 | 4.76 |
| Heisenberg model | [35] | 0.365 | 1.386 | 4.8 |
| 3D Ising model. | [35] | 0.325 | 1.24 | 4.82 |
| 3D XY model | [35] | 0.345 | 1.316 | 4.81 |
| Two-dimensional Ising model. | [36] | 0.125 | 1.75 | 15 |
| Mean field theory | [35] | 0.5 | 1 | 3 |
| Triply critical mean field theory | [37] | 0.25 | 1 | 5 |
| $LaCrSb_3$ | [27] | 0.298 | 1.277 | 5.29 |

This issue can be further clarified by estimating the distance dependence of the magnetic exchange interaction *J*(*r*). For homogeneous magnets, renormalization group theory indicates the decay relationship of *J*(*r*) with *r*:

$$J(r) \sim 1/r^{(d+\sigma)}, \qquad (8)$$

Here, *d* denotes the spatial dimension, and *σ* is a positive constant. Its relationship with *γ* is as follows[38,39]:

$$\gamma = 1 + \frac{4}{d}\frac{n+2}{n+8}\Delta\sigma + \frac{8(n+2)(n-4)}{d^2(n+8)^2} \times [1 + \frac{2G(d/2)(7n+20)}{(n-4)(n+8)}]\Delta\sigma^2, \qquad (9)$$

In the equation, $\Delta\sigma = \sigma - d/2$ and $G(d/2) = 3 - (d/2)^2/4$, where *n* represents the spin degrees of freedom. For magnetic systems with *d* = 3, the exchange interaction *J*(*r*) decays rapidly with distance *r* ($r^{-5}$) when *σ* = 2, which applies to three-dimensional isotropic Heisenberg model magnets. When *σ* = 3/2, *J*(*r*) decays more slowly with distance *r* ($r^{-4.5}$), indicating that the mean-field model holds. By solving Equation (9), we obtain *σ* = 3/2. Consequently, $J(r) \sim r^{-4.96}$, which approaches the prediction of the Heisenberg model. This further confirms the accuracy and reliability of the critical exponents.

Although a certain deviation exists between the experimentally measured critical exponents and the predictions of idealized theoretical models, several analyses collectively demonstrate high consistency and self-consistency. These include Arrott-Noakes analysis, the variation of normalized slope near the critical point, verification of the Widom scaling law, scaling collapse of the magnetic equation of state, and magnetic exchange decay behavior. These results indicate that the magnetic critical behavior of $LaCrSb_3$ near the Curie temperature belongs to the three-dimensional Heisenberg universality class. References [27, 29] fitted the Curie-Weiss law in the high-temperature paramagnetic state and obtained effective magnetic moments per Cr atom of $(3.64 \pm 0.07)\mu_B$ and $3.76\mu_B$, respectively. These values agree well with the theoretical prediction for the high-spin state of $Cr^{3+}$ ions ($S = 3/2$), further supporting the rationality of the spin quantum number $S = 3/2$ in the Heisenberg model. Regarding the lattice dimensionality corresponding to the critical exponents, our results are consistent with Reference [27]. Both studies identify the system as three-dimensional based on the characteristic $\beta > 0.25$.

However, differences exist between this study and Reference [27] regarding the specific classification of the universality class and the mechanism of magnetic interactions. This study explicitly classifies $LaCrSb_3$ into the three-dimensional Heisenberg universality class, with short-range interactions dominating the magnetic interactions in the critical region and paramagnetic state. In contrast, Reference [27] failed to determine the dominant universality class and proposed that long-range interactions dominate the magnetic interactions. We suggest that this discrepancy may stem from differences in crystal growth methods. Crystals grown by the flux method often introduce unintentional doping and additional structural disorder. This reduces the uniformity of the magnetic exchange coupling coefficients, causing the critical exponents to deviate from the characteristics of a fixed universality class. Furthermore, the obtained magnetic exchange interaction decay law $J(r) \sim r^{-4.96}$ (slower than $r^{-5}$) indicates that long-range interactions, such as non-nearest-neighbor interactions and/or Ruderman-Kittel-Kasuya-Yosida (RKKY) interactions, exert a weak perturbative correction effect on the three-dimensional Heisenberg model. This result provides a supplementary explanation for the complexity of the interaction mechanism and the causes of similarities and differences in critical exponents.

Notably, the infinite planar isotropic Heisenberg model typically does not exhibit magnetic order at finite temperatures. However, recent studies on van der Waals bonded quasi-two-dimensional magnetic systems, such as $MPS_3$ ($M$ = Mn, Fe, Ni), $CrXTe_3$ ($X$ = Si, Ge, Te), and $Fe_3GeTe_2$, have confirmed that magnetic interactions in these systems undergo a dimensional crossover from three-dimensional to two-dimensional as temperature decreases[40]. For $LaCrSb_3$, the interlayer bonding is not traditional molecular bonding. It exhibits characteristics similar to typical quasi-two-dimensional magnetic materials like $Fe_3GeTe_2$ in the Deguchi-Takahashi plot. Combined with the synergistic effects of anisotropy evolution and long-range interactions, this makes magnetic order in the two-dimensional limit possible for $LaCrSb_3$. Consequently, its magnetic behavior in the two-dimensional limit holds significant theoretical research value and application prospects.

# 4 Conclusion

We have investigated $LaCrSb_3$, which possesses both quasi-two-dimensional spin fluctuations and three-dimensional magnetic interaction characteristics. Through isothermal magnetization curve tests on single-crystal samples and critical behavior analysis, we clarify the critical characteristics of its ferromagnetic phase transition and the intrinsic magnetic interaction mechanism. Using high-precision isothermal magnetization data near the critical point (115—140 K) and the Arrott-Noakes equation self-consistent iteration method, the Curie temperature is determined to be $T_C$ = 126 K. The critical exponents are derived as $\beta = 0.376$, $\gamma = 1.417$, and $\delta = 4.76$. We verified the reliability of these critical exponents using the Widom scaling law and the magnetic equation of state scaling. Comparison with theoretical models indicates that the critical behavior of the magnetic phase transition in this system belongs to the three-dimensional Heisenberg model universality class. This judgment was further confirmed by the decay relationship of the exchange interaction $J(r)$ with distance $r$, revealing the dominance of isotropic direct exchange interactions in this system. Finally, drawing on research results from other quasi-two-dimensional magnetic materials, we propose that $LaCrSb_3$ may exhibit finite-temperature magnetic order in the two-dimensional limit. Therefore, it possesses significant theoretical research value and practical application prospects.

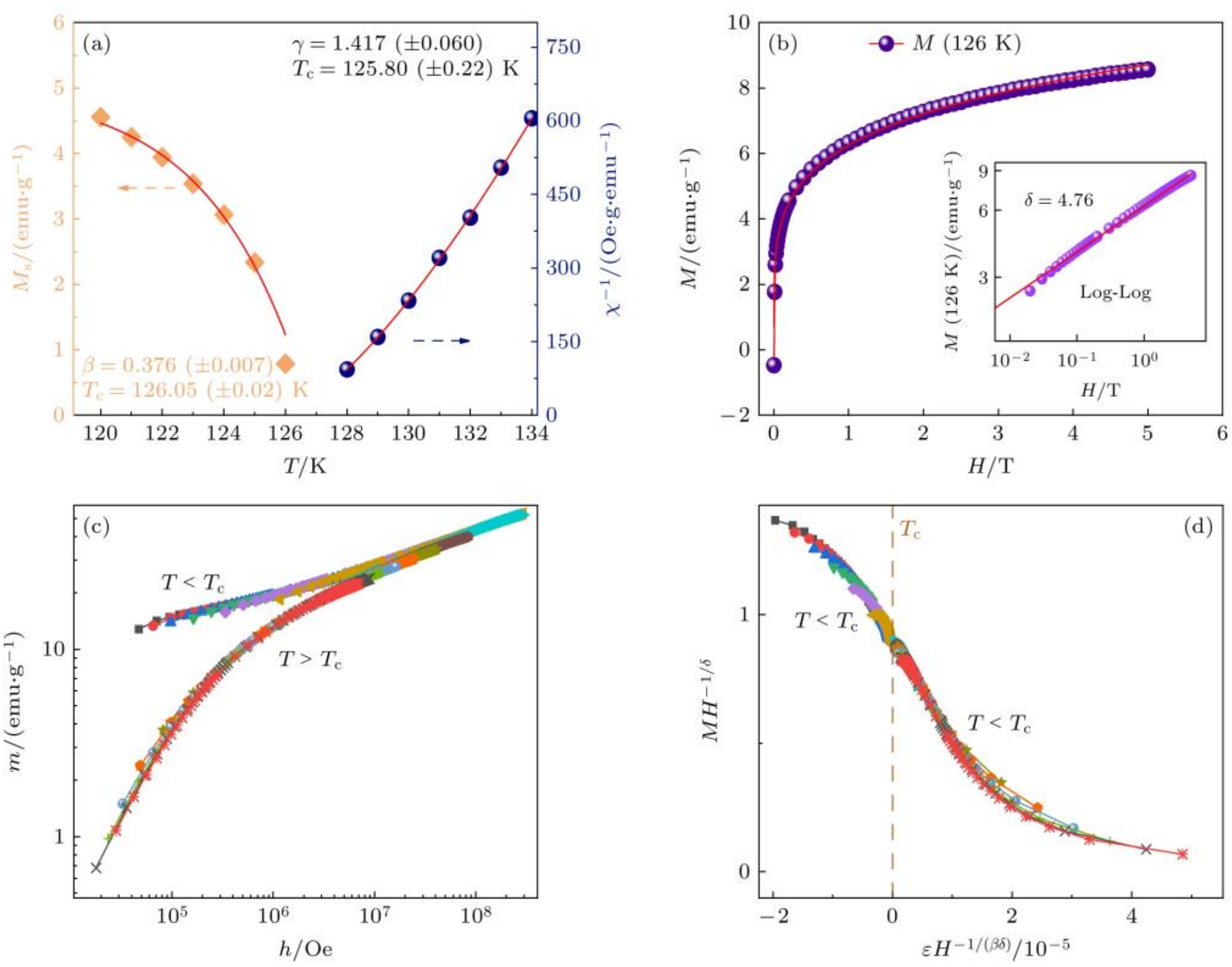